\title{Choosing alpha post hoc: the danger of multiple standard significance thresholds}

\author{Jesse Hemerik\footnote{Econometric Institute, Erasmus University Rotterdam, P.O. Box 1738, 3000 DR Rotterdam, The Netherlands. e-mail: hemerik@ese.eur.nl}\hspace{.1cm} and Nick W. Koning\footnote{Econometric Institute, Erasmus University Rotterdam}}

\documentclass[twocolumn,11pt]{article}

\usepackage{amsmath}
\usepackage{amsfonts}
\usepackage{bm}
\usepackage{bbm}
\usepackage{amsthm}
\usepackage{comment}
\usepackage[nottoc]{tocbibind}
\usepackage{graphicx}

\usepackage{subcaption}

\usepackage{changepage}
\usepackage[numbers]{natbib}
\usepackage{abstract}
\usepackage[colorlinks=true,citecolor=blue,runcolor=blue,linkcolor=blue, pdfborder={0 0 0}]{hyperref}

\usepackage{tikz}
\usepackage{pgfplots}

\usepackage{algorithm}
\usepackage[noend]{algorithmic}

\theoremstyle{plain}

\theoremstyle{definition}

\usepackage[margin=1.2cm]{geometry}

\definecolor{darkblue}{rgb}{0.0, 0.2, 0.6}

\newcommand{\citp}{\cite}
\newcommand{\citt}{\cite}

\begin{document}
\maketitle

\begin{abstract}
\noindent 
A fundamental assumption of classical hypothesis testing is that the significance threshold $\alpha$ is chosen independently from the data.  The validity of confidence intervals likewise relies on choosing $\alpha$ beforehand.  We point out that the independence of $\alpha$ is guaranteed in practice because, in most fields, there exists one standard $\alpha$ that everyone uses -- so that $\alpha$ is automatically independent of everything. However, there have been recent calls to decrease $\alpha$ from $0.05$ to $0.005$. We note that this may lead to multiple accepted standard thresholds  within one scientific field. For example, different journals may require different significance thresholds. As a consequence, some researchers may be tempted to conveniently choose their $\alpha$ based on their p-value. We use examples to illustrate that this severely invalidates hypothesis tests, and mention some potential solutions.

\end{abstract}

\section{Introduction}
\subsection{The call to decrease significance thresholds} \label{seccall}
    Conducting a hypothesis test requires choosing a ``significance level'' or ``significance threshold'', typically denoted by $\alpha$ \citp{van1998asymptotic,lehmann2022testing,hansen2022econometrics}.
    It is assumed to be prespecified, and therefore independent of the p-value \citp{hubbard2003confusion}. 
    Recently, there has been an intense debate in many fields on whether the threshold should be changed from 0.05 to 0.005 \citp{borenstein2001power,ioannidis2005most,banerjee2009hypothesis,mudge2012setting,martin2012quantitative,kim2015significance,ioannidis2016most, szucs2017empirical,benjamin2018redefine,ioannidis2018proposal, lakens2018justify, held2019assessment, di2020statistical,khan2020transforming,aguinis2021reporting, schnog2021urgent, thakur2022potential, fitzpatrick2024impact}.
    In one important paper that stimulated the debate, 72 influential scientists
    propose changing the ``threshold for statistical significance for claims of new discoveries'' to 0.005 \citp{benjamin2018redefine}.

As a consequence of such proposals, scientists may find themselves in a field where there is not one standard threshold, but multiple: some groups of scientists may support $\alpha=0.005$ and some $\alpha=0.05$. This can also translate into the implicit or explicit preferences  of journals, with some journals preferring $0.005$ and others $0.05$.

\subsection{The statistical danger of differing thresholds}
The message of this  paper is that if within one field, there are multiple standard significance thresholds, then researchers may be tempted to first analyze their data and then choose the threshold $\alpha$ that suits them.

For example, suppose a researcher is interested in a certain null hypothesis and plans to compute a p-value for it. 
In her field there are two standard significance thresholds, $0.005$ and $0.05$,  because some researchers or journal editors in her field advocate 0.005 and others advocate 0.05.
After she has  computed the p-value, she wants to share it in some form, e.g. by presenting it at a conference or submitting it to a journal. Then the choice of where to present or where to submit her results will potentially depend on the p-value that she finds.
For example, if she finds $p=0.003$, then she may prefer  a journal that supports  $\alpha=0.005$ over a journal that supports $\alpha=0.05$,  since then her finding is significant at a more stringent threshold.
Consequently, it would be tempting for the researcher not to fix her $\alpha$ before running her experiment, but to  let her $\alpha$ depend on the journal or conference, which she chooses after obtaining her p-value.
 
The statistical issue that arises, is that the researcher's  $\alpha$ is no longer independent of her p-value.
Consequently, the test loses its usual decision-theoretic interpretation, i.e., there is no guarantee anymore that the type I error probability is below $\alpha$ \citp{hubbard2003confusion,grunwald2024beyond,koning2024post}.
Note that this issue is about more than publication bias. It is about the validity of an individual test, from the perspective of the individual researcher:
when she lets  $\alpha$ depend on the data,  she cannot validly interpret her own statistical test anymore.
In theory, a solution would be that the researcher fixes her significance threshold beforehand, but one can easily imagine a researcher who finds $p=0.003$ and then claims her significance level is $\alpha=0.005$, even though she never fixed  $\alpha$ beforehand.

\subsection{ A related discussion: abandoning significance thresholds}

In parallel to the discussion on what the threshold should be, there has been a discussion on whether significance thresholds should be used at all \citp{hubbard2008p,wasserstein2016asa, mcshane2019abandon,amrhein2019scientists}. Indeed, a downside of significance testing in practice is that it may overemphasize a specific p-value threshold rather than interpreting p-values continuously, considering effect sizes, and using other scientifically relevant information \citp{borenstein1994case,greenland2016statistical,thompson2007effect,betensky2019p,scheel2021hypothesis,imbens2021statistical,muff2022rewriting}. For example, the famous ASA statement on p-values \citp{wasserstein2016asa} notes that ``Scientific conclusions and business or policy decisions
should not be based only on whether a p-value passes
a specific threshold." Note that the ASA statement is not against p-values, but against too much focus on p-value thresholds.

Many statisticians agree with the ASA statement's point, but nevertheless we see that in the recent statistical literature, significance testing is still prominent \citp{lehmann2022testing,hansen2022econometrics}. 
One  reason is that from a mathematical point of view, there is often essentially no difference between  a formula for a significance test, a formula for a p-value and a formula for a confidence interval.
For example, confidence intervals can often be directly derived by `inverting' a test.
Another reason is that some statisticians still value type I error control, which is only possible if some $\alpha$ is chosen.

Beyond theoretical statistics, in science in general 
we see that significance testing has not disappeared \citp{hubbard2019will,mayo2022statistical,thakur2022potential,maier2022justify,hansen2022econometrics,fitzpatrick2024impact}.
Moreover, if one abandons significance tests, then one is still likely to compute confidence intervals, which also require choosing $\alpha$.


\subsection{This paper}
 Throughout the paper, we stay within the classical Neyman-Pearson theory of hypothesis testing. 
Our main point is that tests are often invalidated if $\alpha$ depends on the p-value \citp{hubbard2003confusion}, and that this will likely occur when within a field there are multiple standard thresholds.
 \citt{uygun2023epistemic} make the following related point: one possible reason why a single $\alpha$ has been the norm for so long, is that it prevents researchers from choosing $\alpha$ post hoc. 

Further, we discuss in detail what can go wrong in terms of type I error control.
We evaluate what the type I error rate is, conditional on a certain $\alpha$ being chosen based on the p-value.
Moreover, we introduce a measure of how problematic a test with  data-dependent  $\alpha$ is.
If this measure is large, this means the conditional type I error rate of the test is  too large in an average sense.
Apart from discussing the situation where the researcher chooses between two significance levels, we will also consider the most extreme scenario where there are many choices for $\alpha$. 
In Section \ref{sece} and the Discussion, we briefly point out some potential solutions to the discussed issue.

\section{Discrepancies between $\alpha$ and the type I error rate}   \label{seconehypo}

\subsection{Notation and concepts} \label{secgeneral}
Consider a null hypothesis $H_0$ and a corresponding  test $\phi_\alpha$, where $0< \alpha \leq 1$ indicates the chosen significance level.  The test $\phi_\alpha=\phi_\alpha(X)$ is a function of the data $X$ and maps to the set $\{0,1\}$, where $1$ indicates that $H_0$ is rejected.
In this paper, all mentioned probabilities and expectations will be under $H_0$.

Suppose that $\alpha$ possibly depends on $X$. Let $A\subseteq[0,1]$ denote the set of possible values that $\alpha$ can take. Throughout we use `$a$' to denote a value from $A$, independent of the data. We can then ask whether conditional on the data-dependent $\alpha$ taking the value $a$, the type I error rate is below $\alpha$, i.e. whether  
$\mathbb{P}(\phi_\alpha=1\mid\alpha=a)\leq a$. 
A measure of the discrepancy between $a$ and $\mathbb{P}(\phi_\alpha=1\mid\alpha=a)$ is the difference
\begin{equation*} \label{defdiscr}
d_a:=\mathbb{P}\big(\phi_\alpha=1\mid\alpha=a\big)-a.
\end{equation*}
Ideally, we would have $d_a=0$ for all $a\in A$, but this is often not the case for many $a\in A$, if $\alpha$ depends on the data. This is well known \citp{hubbard2003confusion}  and   illustrated in Sections \ref{seca1a2} and  \ref{secmanya}.

Now suppose that for some $a\in A$, the difference is $d_a = 0.01$. Do we consider this difference small or large? To a large extent, this depends on what $a$ is. For example, if $a=0.05$, then having $d_a = 0.01$ is not highly problematic, since  $\mathbb{P}(\phi_\alpha=1\mid\alpha=a)=0.06$ is then not much larger than $a$, in relative terms. However, if $a=0.005$, then $d_a = 0.01$ implies that the true error rate $\mathbb{P}(\phi_\alpha=1\mid\alpha=a)=0.015$ is three times larger than $a$. In that case, $d_a = 0.01$ is a huge difference. Thus, $d_a$ is arguably not the best measure of how liberal  the test is.

Instead it can be more  meaningful to look at the \emph{relative} discrepancy
\begin{equation*} \label{eqratio}
r_a:=\mathbb{P}\big(\phi_\alpha=1\mid\alpha=a\big)/a. 
\end{equation*}
We will call $r_a$ the \emph{discrepancy ratio conditional on} $\alpha=a$.  We would like this ratio to be at most 1. If  $r_a>1$, the test is liberal conditional on $\alpha=a$.
Note that for a given number $a$, $r_a$ is a fixed number. In contrast, since $\alpha$ depends on the data, $r_\alpha$ is data-dependent.

If the discrepancy ratio $r_a$ is below 1 for every $a\in A$, then in particular $\mathbb{E} r_{\alpha}$  will be below 1.
If on the other hand $\mathbb{E} r_{\alpha}$ is much larger than $1$ under $H_0$, it follows that either there is at least a small probability that $r_{\alpha}$ is much larger than $1$, or there is a large probability that $r_{\alpha}$ is at least somewhat larger than 1. Both these possibilities are problematic for researchers. 
Thus, $\mathbb{E} r_{\alpha}$ is a summary measure of how wrong things go on average.
If $\mathbb{E} r_{\alpha}\gg 1$, this  indicates that the test procedure is seriously flawed, in an average sense.

It is useful to note that   $\mathbb{E} r_{\alpha}$ can be rewritten as  a simple expectation:
\begin{align}
    \mathbb{E} r_{\alpha} 
        &= \mathbb{E}\big\{ \mathbb{P} (\phi_\alpha=1\mid\alpha)/\alpha \big\} 
        = \mathbb{E}\big\{   \mathbb{E} \big( \phi_\alpha |\alpha\big)/\alpha\big\}  \nonumber \\
        &= \mathbb{E}\big\{   \mathbb{E} \big(\phi_\alpha/\alpha \mid\alpha\big)\big\} 
        =  \mathbb{E} \big(\phi_\alpha/\alpha\big).  \label{eqrewrite} 
\end{align}

\subsection{Example 1: two possible significance levels $\alpha$} \label{seca1a2}
Consider a researcher who tests one hypothesis. 
She computes a p-value, which we assume  to exactly satisfy
\begin{align*}
    \mathbb{P}(p \leq \alpha) = \alpha
\end{align*}
under $H_0$ for every data-independent $\alpha$.
Suppose that in her field there are two standard significance thresholds, $a_1$ and $a_2$, where $0<a_1<a_2<1$.
As a simple behavioral model, assume she chooses the smaller threshold $\alpha=a_1$ if $p\leq a_1$ and chooses $\alpha=a_2$ if $ a_1<p\leq 1$, so that $\alpha $ depends on $p$ and hence on the data.

What happens to the type I error probability conditional on $\alpha=a_1$ can be seen immediately: if $\alpha=a_1$, then this means that $p$ was apparently at most $a_1$, which means that $H_0$ is rejected. Thus, under $H_0$, conditional on $\alpha=a_1$, the type I error probability is not $a_1$ but 1. 
Conditional on $\alpha = a_2$, the type I error probability is strictly smaller than $a_2$,
\begin{align*}
    \mathbb{P}(p \leq \alpha \mid \alpha = a_2)
        &= \mathbb{P}(p \leq a_2 \mid p > a_1) \\
        &= \frac{a_2 - a_1}{1 - a_1} < \frac{a_2 - a_2a_1}{1 - a_1}
        = a_2.
\end{align*}
Thus, the discrepancy ratio is $r_{\alpha}>1$ if $\alpha=a_1$ and  $r_{\alpha}<1$ if  $\alpha=a_2$.

Next, we can ask whether perhaps we control the type I error probability in expectation, in the sense that $\mathbb{E}r_{\alpha}\leq 1$ under $H_0$. 
This turns out to not be the case either, since under $H_0$, by \eqref{eqrewrite},
\begin{align*}
    \mathbb{E}r_\alpha
        &= \mathbb{E}(\phi_\alpha / \alpha) \\
        &= \mathbb{P}(p\leq a_1)   \mathbb{E}(\phi_{\alpha}/\alpha\mid p\leq a_1) \\
        &+ \mathbb{P}(a_1<p \leq a_2)   \mathbb{E}(\phi_{\alpha}/\alpha\mid a_1<p \leq a_2) \\
        &= a_1(1/a_1) + (a_2-a_1)(1/a_2) \\
        &= 1+(a_2-a_1)/a_2 > 1.
\end{align*}
For example, if $a_1=0.005$ and $a_2=0.05$, then $\mathbb{E}r_{\alpha}=1.9$.  We conclude that both conditionally and on average, the researcher's test does not provide type I error control.

\subsection{Example 2: Infinitely many significance levels $\alpha$} \label{secmanya}
Now suppose there are not just two, but many thresholds to choose from.
In fact, let us take this to the extreme by allowing the researcher to choose any threshold $0<\alpha \leq C$ up to some constant $C > 0$, e.g. $C = 0.05$ or $C = 0.1$.
If the researcher desires to report a discovery, it is easy to imagine that she is biased towards choosing an $\alpha$ that is at least as large as her p-value $p$, when possible.
Again taking the most extreme choice, suppose she selects $\alpha=p$, unless $p>C$, in which case she selects $\alpha=C$.

In case $\alpha<C$, $H_0$ is always rejected, so that we do not have conditional type I error rate control.
Indeed, for every $a<C$, we have $r_a= 1/a > 1$. 
 Next, we may ask whether we have type I error control not conditionally, but at least on average, in the sense that
$\mathbb{E}r_{\alpha}\leq 1$ under $H_0$.
Since $p$ is uniform on $[0,1]$ and  $p\leq \alpha$ if and only if $p\leq C$, we have 
\begin{align*}
    \mathbb{E}r_{\alpha}
        &= \mathbb{E} (\phi_{\alpha}/\alpha) \\
        &= \mathbb{P}( p> \alpha)\cdot 0+ \mathbb{P}( p\leq \alpha) \mathbb{E} (\phi_{\alpha}/\alpha \mid p\leq \alpha) \\
        &= \mathbb{P}( p\leq \alpha) \mathbb{E} (\phi_{\alpha}/\alpha \mid\alpha=p\leq C) \\
        &= \mathbb{P}( p\leq \alpha) \mathbb{E} (1/p \mid p\leq C) \\
        &= CC^{-1}\int_0^C x^{-1} dx =\log(x) \Big|_0^C = \infty.
\end{align*}
We see that the expected discrepancy ratio $\mathbb{E}r_{\alpha}$ is not only larger than 1, but in fact infinity.
Thus, in expectation, the test is completely out of control.

\section{Connection to e-values} \label{sece}
    In case we are only interested in $\mathbb{E}r_{\alpha}$, then a solution is offered by the e-value.
    An e-value is a measure of evidence that has been recently proposed as an alternative to the p-value \citp{howard2021time, shafer2021testing, vovk2021values, grunwald2023safe, ramdas2023gametheoretic, koning2024measuring}.
    An e-value is typically defined as a non-negative random variable $e$, whose expectation is bounded by 1 under $H_0$:
    \begin{align*}
        \mathbb{E}\, e \leq 1.
    \end{align*}
    In the context of a simple null and alternative hypothesis that only contain a single distribution, a prominent example of an $e$-value is the likelihood ratio between these distributions \citp{shafer2021testing}.
    For a general null, desirable e-values are increasing functions of the likelihood ratio between the alternative and a kind of least-favorable distribution in the null \citp{grunwald2023safe, larsson2024numeraire, koning2024measuring}.

    The reciprocal $p^* = 1/e$ of an e-value gives rise to a conservative p-value.
    Indeed, every such $p$-value is valid:
    \begin{align*}
        \mathbb{P}(p^* \leq \alpha)
             = \mathbb{P}(e \geq 1/\alpha) 
             \leq \alpha\,\mathbb{E}\,e
             \leq \alpha,
    \end{align*}
    for all data-independent $\alpha > 0$, where the first inequality follows from Markov's inequality and the second inequality from the definition of the e-value.

    Recently, \citp{koning2024post} showed that $p^*$ is not just some overly conservative p-value, but exactly the kind of p-value that 
    satisfies
    \begin{align*} \label{ratiobelow1}
        \mathbb{E}\, r_\alpha
            = \mathbb{E}\left(\frac{\mathbb{P}(p^* \leq \alpha\mid \alpha)}{\alpha}\right) \leq 1
    \end{align*}
    regardless of how $\alpha$ depends on the data.
    This is in stark contrast to traditional p-values, for which $\mathbb{E}r_{\alpha}$ can be $\infty$, as discussed in Section \ref{secmanya}.

\color{black}

\section*{Discussion}
The fundamental point of this paper is that significance testing works in practice because researchers all by default use the same $\alpha$.
This ensures that $\alpha$ is independent of the data \citp{uygun2023epistemic}. 
If there are multiple standards for $\alpha$ in a field and a researcher chooses her $\alpha$ post hoc, then that  spells the end of the validity of her  tests.  Likewise, confidence intervals are no longer valid if $\alpha$ is chosen post hoc. 
One solution would be that  researchers  fix $\alpha$ in advance and do not change it, but they would not necessarily have enough incentives or awareness to do this.

When a community of researchers within a scientific field starts adopting a lower $\alpha$, say $\alpha=0.005$, there are two possible scenarios: 1. Everyone in that field adopts $\alpha=0.005$ at the same time; 2. some stick to $\alpha=0.05$ while others adopt $\alpha=0.005$.
The problem described in this paper occurs in scenario 2.

Existing papers that advocate decreasing $\alpha$  \citp{johnson2017reproducibility,szucs2017empirical,benjamin2018redefine,ioannidis2018proposal,held2019assessment, schnog2021urgent, fitzpatrick2024impact},  do not discuss the following question: would they already consider it an improvement if \emph{some} people or journals in a field decrease $\alpha$, or do they consider  it essential that the field as a whole uses the same  $\alpha$? The present paper illustrates that these distinctions matter.

Suppose that in a  field the scenario occurs where some stick to $\alpha=0.05$ while others adopt $\alpha=0.005$---thus  providing an incentive for letting $\alpha$ depend on the data. One solution would be  to require pre-registration of  $\alpha$ before researchers conduct an experiment \citp{maier2022justify}.
Another partial solution would be to use e-values instead of p-values, as discussed in Section  \ref{sece}.
Another way out is that a researcher uses a Bayesian approach, which does not require fixing an $\alpha$ and nevertheless tells us the probabilty that $H_0$ is false -- which frequentist statistics never does \citp{kelter2020bayesian,lakens2020improving,van2021bayesian}. A potential downside of Bayesian inference is that one needs to specify a prior distribution, which often depends on subjective considerations.

\setlength{\bibsep}{3pt plus 0.3ex}  
\def\bibfont{\small}  

\bibliographystyle{biblstyle}
\bibliography{bibliography}

\end{document}